\documentclass[conference]{IEEEtran}
\IEEEoverridecommandlockouts
\usepackage{cite}
\usepackage{amsmath,amssymb,amsfonts}
\usepackage{array}
\usepackage[caption=false,font=footnotesize]{subfig}
\usepackage{textcomp}
\usepackage{url}
\usepackage{verbatim}
\usepackage{graphicx}
\usepackage[colorlinks=true, linkcolor=blue, citecolor=blue, urlcolor=blue]{hyperref}
\usepackage{makecell} 
\usepackage{enumitem}

\usepackage[linesnumbered,ruled,boxed]{algorithm2e}
\renewcommand{\KwData}{\textbf{Input: }}  
\renewcommand{\KwResult}{\textbf{Output: }} 
\usepackage{xcolor}
\usepackage{fancyhdr}

\fancypagestyle{wfiot}{%
	\fancyhf{}
	\fancyhead[L]{\footnotesize\leftmark}
	\fancyhead[R]{\footnotesize\thepage}
	
}

\makeatletter

\def\ps@IEEEtitlepagestyle{%
	\def\@oddhead{%
		\hfil
		\footnotesize
		\leftmark
		\hfil
	}%
	\def\@evenhead{%
		\hfil
		\footnotesize
		\leftmark
		\hfil
	}%
	\def\@oddfoot{%
		\parbox{\textwidth}{%
			\centering
			\scriptsize
			\copyright{} 2026 IEEE.
			Personal use of this material is permitted.
			Permission from IEEE must be obtained for all other uses,
			in any current or future media, including reprinting/republishing
			this material for advertising or promotional purposes, creating new
			collective works, for resale or redistribution to servers or lists,
			or reuse of any copyrighted component of this work in other works.
		}%
	}%
	\def\@evenfoot{%
		\@oddfoot
	}%
}

\makeatother

\begin{document}
	
	
	\title{When Time Meets Space: Entropy Integration and Dynamic Threshold for Adaptive DDoS Detection in SDN}

	\author{
		\IEEEauthorblockN{
			Zhaoyang Zhang\IEEEauthorrefmark{1},
			Shen Wang\IEEEauthorrefmark{1},
			Ahmad Taha\IEEEauthorrefmark{2},
			Elias Bou-Harb\IEEEauthorrefmark{3}, and
			Xiaofeng Tao\IEEEauthorrefmark{1}}
		\IEEEauthorblockA{
			\IEEEauthorrefmark{1}National Engineering Research Center of Mobile Network Technologies,\\
			Beijing University of Posts and Telecommunications, Beijing 100876, China;
			{\small \{zhangzhaoyang, shen.wang, taoxf\}@bupt.edu.cn}\\
			\IEEEauthorrefmark{2}Departments of Civil and Environmental Engineering and Electrical and Computer Engineering,\\
			Vanderbilt University, Nashville, TN, USA; ahmad.taha@vanderbilt.edu\\
			\IEEEauthorrefmark{3}Division of Computer Science \& Engineering,\\
			Louisiana State University, Baton Rouge, LA, USA; ebouharb@lsu.edu}
		\thanks{Corresponding author: Shen Wang. This work is supported by the NSFC under grants No. 62203062 and No. 61932005, the Fundamental Research Funds for the Central Universities under Grant 2242022k60006, and NSF under grant No. 2152450.
		}
	}

	\maketitle
	
	\markboth
	  {To Appear in 12th World Forum on Internet of Things,
	   Abu Dhabi, United Arab Emirates, 26--29 October, 2026}
	  {}

	\begin{abstract}

        Entropy-based Distributed Denial of Service (DDoS) detection in Software-Defined Networking (SDN) commonly relies on spatial traffic distributions and static or loosely adaptive thresholds, making it vulnerable to legitimate traffic fluctuations in Internet of Things (IoT) environments. This paper proposes a lightweight spatiotemporal entropy-based detector for DDoS attacks. Spatial entropy is computed from dynamically selected traffic attribute pairs, while temporal entropy captures the randomness of packet inter-arrival times. The two normalized entropy measures are fused into a unified indicator and evaluated using a constrained second-order Exponentially Weighted Moving Average threshold that jointly tracks entropy trend and volatility. To prevent attack-contaminated observations from biasing threshold adaptation, threshold updates are performed only for windows classified as normal. Testbed results show 99.26\% recall, a 0.9737 F1-score, and a 3.2\% false positive rate (41.74\% below that of spatial entropy alone). On CICDDoS2019, the method achieves an FPR of 0 and remains competitive with machine-learning-based methods. It requires 3.95 ms of core processing per window and 11.65\% system-wide CPU utilization, supporting resource-constrained edge and IoT deployment.
		
	\end{abstract}
	
	
	\section{Introduction and Paper Contributions}\label{sec:intro}
	
	Software-Defined Networking (SDN) provides centralized control and monitoring but also exposes controllers and control-plane resources to Distributed Denial of Service (DDoS) attacks, making them critical targets in Internet of Things (IoT) and edge environments. Malicious traffic may exhaust links, flow tables, or controller processing resources and degrade legitimate services~\cite{chahal2024ddos}.

	Entropy-based detection is widely used due to its lightweight and interpretable nature~\cite{Conditional_Entropy_2019,JESS2018}. Existing methods typically rely on spatial traffic distributions (e.g., IPs, ports, protocols), where attack concentration reduces entropy. However, benign traffic may also exhibit short-term concentration, while fixed thresholds fail to adapt to dynamic and heterogeneous IoT workloads.
	
\begin{figure}
		\centering
		\includegraphics[width=3in]{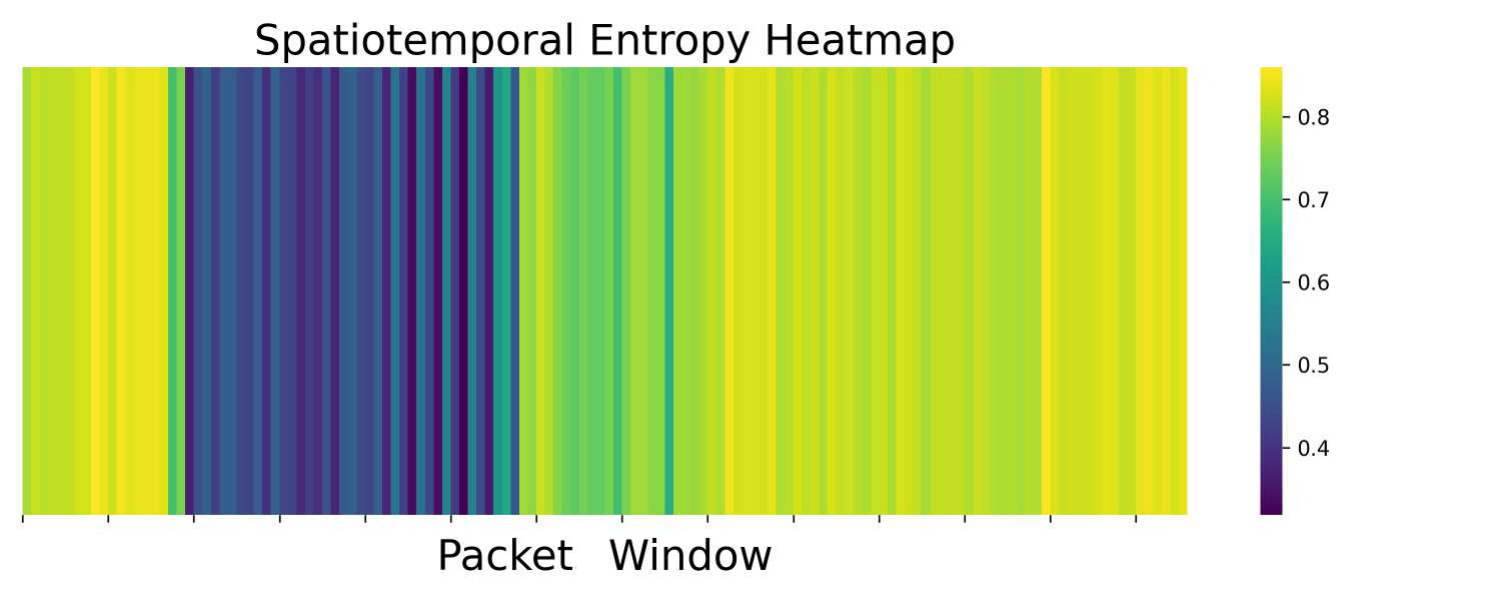}
		\caption{Spatiotemporal entropy patterns; dark blue denotes DDoS traffic.}
		\label{fig:Spatiotemporal_entropy_heatmap}
	\end{figure}

	To overcome these limitations, we propose a spatiotemporal detection framework that fuses spatial entropy with temporal entropy (i.e., packet inter-arrival times) and employs a second-order Exponentially Weighted Moving Average (EWMA) threshold. Spatial entropy captures attack aggregation, temporal entropy distinguishes automated flooding from transient legitimate concentration, and constrained adaptation prevents attacks from entering the baseline. Figure~\ref{fig:Spatiotemporal_entropy_heatmap} illustrates the separability of the resulting representation.
	
	\subsection{Related Work and Research Gap}\label{subsec:related_works}
	Statistical methods primarily rely on traffic feature statistics to identify anomalies, with entropy being a widely adopted indicator. Conditional entropy provides lightweight SDN detection~\cite{Conditional_Entropy_2019}, while joint entropy leverages multiple features to mitigate both known and previously unseen attacks~\cite{JESS2018}. However, these methods primarily employ spatial entropy metrics, and their omission of temporal patterns limits detection robustness. Arshadi \textit{et al.} used the entropy of SYN packet arrival intervals as a randomness indicator for offline and real-time detection~\cite{time_entropy_SYN_2011}. Temporal entropy loses sensitivity when attack timing is diverse, so the two dimensions are complementary rather than interchangeable.
	
	Statistical methods compare entropy with a decision threshold. Fixed thresholds are difficult to apply under normal traffic fluctuations, motivating dynamic mechanisms based on sliding windows, dispersion, or EWMA updates~\cite{Entropy_2022_TNSM,P4_2022_TDSC}. However, indiscriminate updates may absorb attacks and lose sensitivity; both the current trend and short-term variability of normal traffic must be represented.
	
	Machine-learning and hybrid detectors achieve high accuracy~\cite{LSTM_2022_TCCN,Entropy+SVM_2022_Computers&Security,Entropy+ML_2024_SciRep}, but rely heavily on labeled data and computing resources, and deep models generally lack interpretability. We instead combine complementary entropy evidence with a threshold updated only by accepted normal windows.
	
	\subsection{Paper Contributions and Organization}
	
	The main contributions are as follows:
	\begin{enumerate}[label=(\roman*)] 
		\item We design a normalized spatiotemporal entropy representation that jointly models structural (spatial) and temporal traffic uncertainty, providing a unified lightweight indicator for SDN DDoS detection without requiring training data.
		\item We develop a constrained second-order EWMA state evolution mechanism that jointly tracks entropy trend and volatility, while preventing attack-contaminated observations from updating the statistical state, thereby improving stability under non-stationary traffic.
		\item Testbed averages show 99.26\% recall and a 3.2\% false positive rate (41.74\% below that of spatial entropy alone), with 3.95 ms of core processing per window and 11.65\% system-wide CPU utilization. The method also remains competitive on CICDDoS2019 without model training.
	\end{enumerate}
	
	Sections~\ref{sec:method} and~\ref{sec:dynamic_threshold} present the detection and adaptive threshold design; Section~\ref{sec:experiment} evaluates the method; and Section~\ref{sec:conclusion} concludes the paper.

	\section{Spatiotemporal Entropy Detection Scheme via Dynamic Thresholds}\label{sec:method}
	
	\subsection{Threat Assumptions}

	We consider volumetric DDoS attacks generated by distributed hosts,
	including TCP SYN, TCP ACK, and UDP flooding traffic.
	The SDN controller is assumed to be trusted and continuously collects
	packet attributes and timestamps from switches.
	The proposed detector operates in the control plane without requiring
	data-plane modifications.
	Controller compromise, physical tampering, and stealthy low-rate attacks
	that closely mimic legitimate traffic are outside the scope of this work.
	
	\subsection{SDN-Based DDoS Detection Framework}\label{sec:framework}
	
\begin{figure*}
		\centering
		\includegraphics[width=0.9\textwidth]{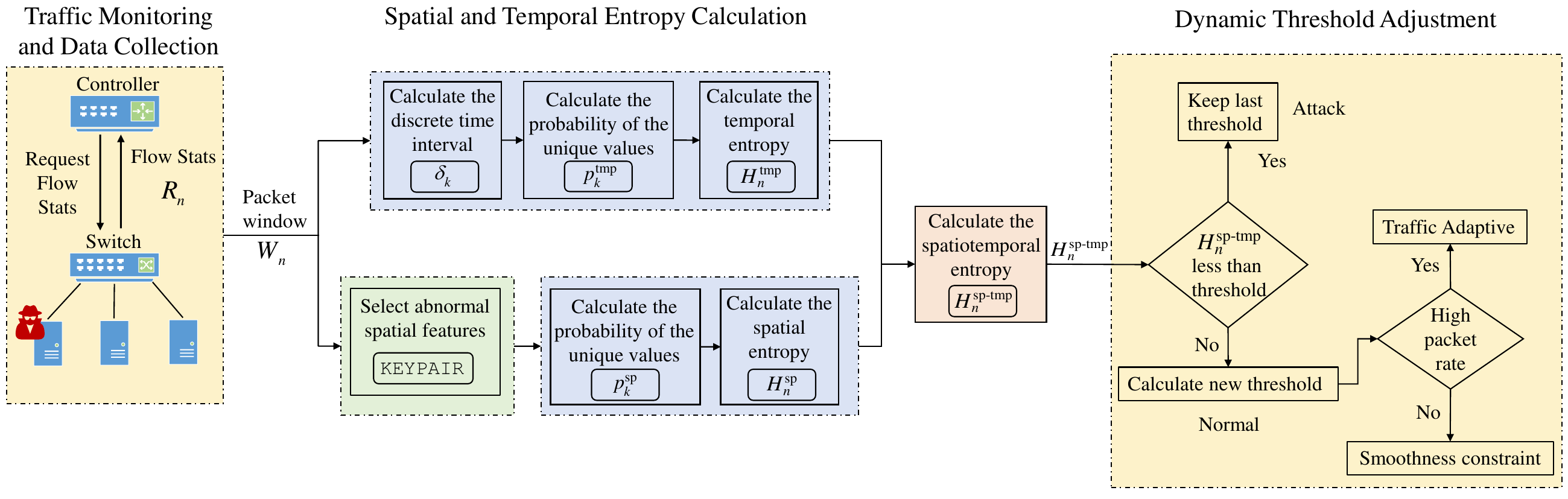}
		\caption{SDN DDoS detection through traffic collection, key-pair selection, spatiotemporal entropy integration, and adaptive EWMA thresholding.}
		\label{fig:framework}
		\vspace{-1.2em}
	\end{figure*}

	As shown in Figure~\ref{fig:framework}, the controller maintains a packet
	window $W_n=\{d_1,\ldots,d_N\}$ and continuously extracts traffic features
	for spatiotemporal entropy analysis and adaptive-threshold classification. Among the eight traffic attributes (i.e., $IP_{src}$, $IP_{dst}$, $TTL$, $PORT_{src}$, $PORT_{dst}$, $Protocol$, $PKT_{len}$, and $TCP_{flag}$),
	28 pairwise combinations can be formed.
	Following~\cite{JESS2018}, the controller compares the distribution of
	each pair with a normal profile derived from the MAWI traffic archive
	and selects the pair exhibiting the largest deviation as
	\texttt{KEYPAIR} $(a_i,a_j)$.
	For example, if an attack concentrates on a specific destination and
	protocol type, $(IP_{dst},Protocol)$ may be selected.
	
	For each window, the controller computes spatial entropy
	$H_n^{\mathrm{sp}}$ from the selected \texttt{KEYPAIR} and temporal
	entropy $H_n^{\mathrm{tmp}}$ from packet inter-arrival times.
	After normalization and fusion, the resulting spatiotemporal entropy
	$H_n^{\mathrm{sp\text{-}tmp}}$ is compared with the adaptive threshold.
	A detected attack freezes threshold updates, whereas attack-free windows
	are used to update the EWMA statistics.

	\subsection{Spatiotemporal Entropy Integration}\label{subsec:spatiotemporal_entropy}
	Spatial entropy measures selected-attribute concentration, whereas temporal entropy measures inter-arrival-time diversity; both are normalized to $[0,1]$.

	\subsubsection{Spatial Entropy}\label{subsubsec:spatial_entropy}
	
	Spatial entropy measures the concentration of traffic over selected attribute-value pairs. Attack traffic often targets a limited set of destinations, ports, or protocols, producing a skewed distribution, whereas normal traffic is typically more diverse. For each packet $d_l$ in window $W_n$, the corresponding attribute-value pair is $ (x_l,y_l)=\bigl(d_l[a_i],d_l[a_j]\bigr), $ where $d_l[a_i]$ and $d_l[a_j]$ denote the values of attributes $a_i$ and $a_j$ carried by packet $d_l$. Collecting all packets in the window yields the multiset $ \mathcal{M}_n^{\mathrm{sp}} = \{(x_1,y_1),\ldots,(x_N,y_N)\}. $ Let $K^{\mathrm{sp}}$ denote the number of distinct attribute-value pairs in $\mathcal{M}_n^{\mathrm{sp}}$. For a distinct pair $(x_k,y_k)$, let $m_n(x_k,y_k)$ denote its occurrence count within the window. The empirical probability distribution is  
	\begin{equation}
		p_k^{\mathrm{sp}}
		=
		\frac{m_n(x_k,y_k)}{N},
		\label{eq:p_sp}
	\end{equation}
	where $N=|\mathcal{M}_n^{\mathrm{sp}}|$ is the number of packets in the
	window.
	
		The normalized Shannon entropy~\cite{shannon1948mathematical} is
	\begin{equation}
		H^{\mathrm{sp}}_n =
		\begin{cases}
			0, & K^{\mathrm{sp}}=1, \\
			-\dfrac{\sum_{k=1}^{K^{\mathrm{sp}}}p_k^{\mathrm{sp}}\log_2p_k^{\mathrm{sp}}}{\log_2K^{\mathrm{sp}}}, & K^{\mathrm{sp}}>1.
		\end{cases}
		\label{eq:spatial_entropy_multiset}
	\end{equation}
	  Normalization by $\log_2K^{\mathrm{sp}}$ ensures
$H_n^{\mathrm{sp}}\in[0,1]$, where lower values indicate stronger concentration of traffic over a small number of attribute-value pairs.

	\subsubsection{Temporal Entropy}\label{subsubsec:time_entropy}
	Temporal entropy captures the randomness of packet inter-arrival times rather than traffic attributes. DDoS traffic is commonly generated by botnets sending large numbers of requests at fixed or near-fixed rates, producing concentrated intervals. In contrast, normal activities such as user browsing exhibit greater randomness in packet timing. This difference makes temporal entropy useful for distinguishing normal and malicious traffic independently of packet fields.
	
\begin{figure}[!t]
		\centering
		\includegraphics[width=0.8\linewidth]{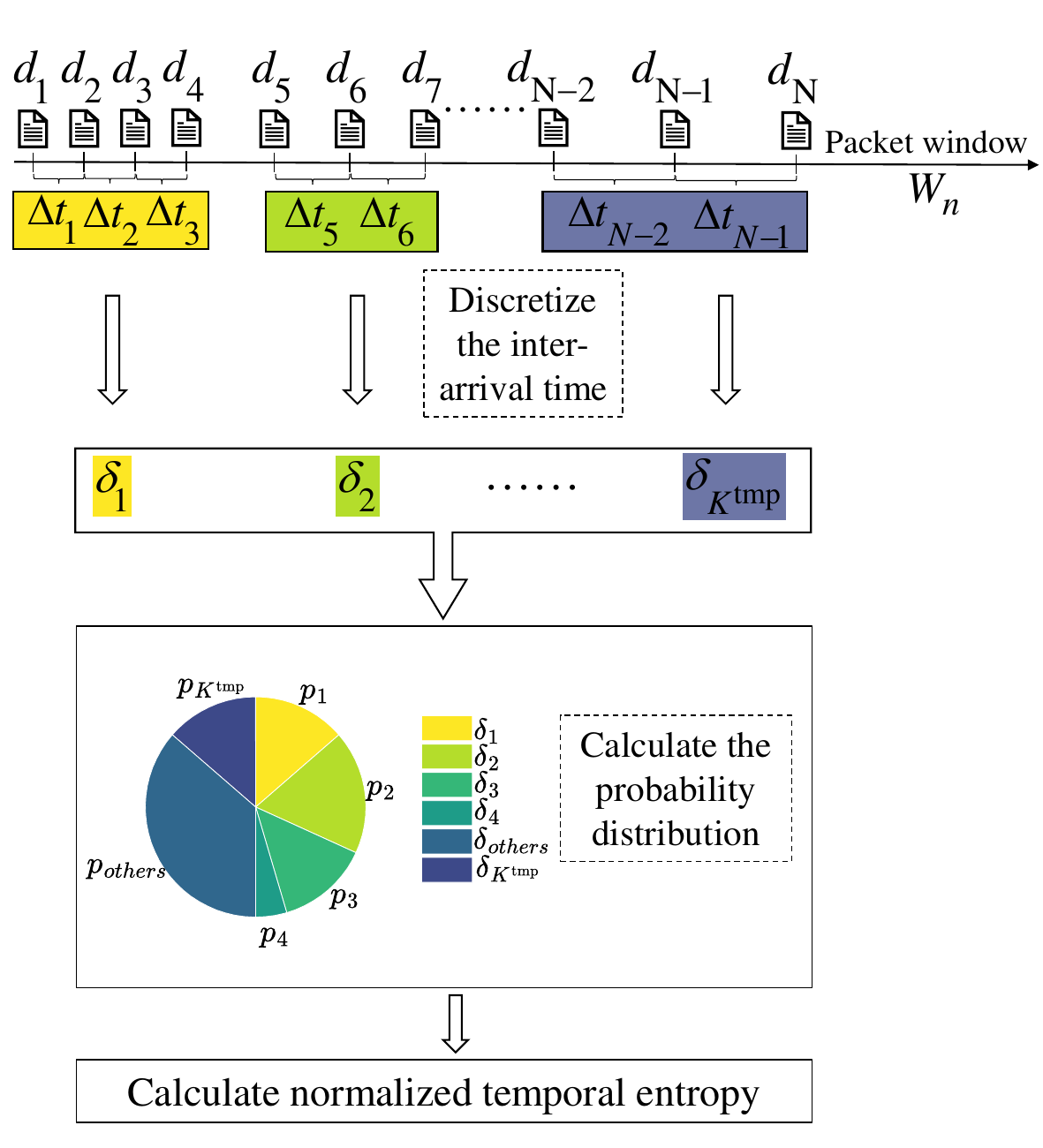}
		\caption{Temporal entropy from quantized inter-arrival times.}
		\label{fig:Temporal_entropy_diagram}
	\end{figure}

	For $\Delta t_i=t_{i+1}-t_i$, we quantize $\Delta t'_i=\Delta_b\lfloor\Delta t_i/\Delta_b+1/2\rfloor$ and form $\mathcal{M}_n^{\mathrm{tmp}}=\{\Delta t'_1,\ldots,\Delta t'_{N-1}\}$, as illustrated in Figure~\ref{fig:Temporal_entropy_diagram}. A smaller $\Delta_b$ preserves finer details but is more susceptible to clock precision and network jitter, whereas a larger value improves noise robustness at the cost of potentially masking attack signatures. The discretization groups small timing variations and improves the robustness of entropy estimation.
	
	The support set is $\operatorname{supp}(\mathcal{M}_n^{\mathrm{tmp}})=\{\delta_1,\ldots,\delta_{K^{\mathrm{tmp}}}\}$, where each $\delta_k$ is a distinct discretized inter-arrival-time value and $K^{\mathrm{tmp}}$ is the number of such values. With occurrence count $m_n(\delta_k)$, we have
	\begin{equation}
		p_k^{\mathrm{tmp}} = \frac{m_n(\delta_k)}{N-1},
		\label{eq:p_tmp}
	\end{equation}
	\begin{equation}
		H^{\mathrm{tmp}}_n =
		\begin{cases}
			0, & K^{\mathrm{tmp}}=1, \\
			-\dfrac{\sum_{k=1}^{K^{\mathrm{tmp}}}p_k^{\mathrm{tmp}}\log_2p_k^{\mathrm{tmp}}}{\log_2K^{\mathrm{tmp}}}, & K^{\mathrm{tmp}}>1.
		\end{cases}
		\label{eq:temporal_entropy_multiset}
	\end{equation}
	
	Lower $H_n^{\mathrm{tmp}}$ indicates a concentrated interval distribution in which packets arrive at fixed or near-constant intervals, as commonly observed in botnet-driven flooding. Higher entropy reflects more random packet arrivals typical of legitimate interactive behavior. Nevertheless, rate limiting, multiple bots, or network jitter may diversify malicious intervals, so temporal entropy alone cannot detect every attack.

	\subsubsection{Spatiotemporal Entropy}\label{subsubsec:Spatiotemporal_entropy}
	
	The two normalized components describe different evidence and are fused as
	\begin{equation}
		H^\mathrm{sp\text{-}tmp}_n = \alpha H^\mathrm{{sp}}_n + (1-\alpha)H^\mathrm{tmp}_n,
		\label{eq:spatiotemporal_entropy} 
	\end{equation}
	where $\alpha\in(0,1)$ determines the relative weighting of spatial and temporal entropy. The convex combination preserves interpretability and keeps the fused value normalized because both components are bounded. It also provides a tunable trade-off without nonlinear parameters or additional training and produces a scalar feature suitable for dynamic thresholding. We use $\alpha=0.5$; Figure~\ref{fig:alpha} shows stable performance for $\alpha\in[0.3,0.8]$.
	
	\subsubsection{Spatiotemporal Entropy Computation Algorithm}\label{subsubsec:entropy_algorithm}
	
	Algorithm~\ref{alg:entropy_computation} counts selected spatial pairs and discretized inter-arrival times, then fuses their normalized entropies in one window scan without training data.

	\begin{algorithm}
		\caption{Spatiotemporal Entropy Computation}
		\label{alg:entropy_computation}
		\small \DontPrintSemicolon
		\KwData{Packet window $W_n = \{d_1, d_2, \dots, d_N\}$, \texttt{KEYPAIR} $(a_i, a_j)$, time discretization parameter $\Delta_b$}, balancing weight $\alpha \in (0,1)$ \;
		\KwResult{Spatiotemporal entropy $H^\mathrm{sp\text{-}tmp}_n$}
		
		\textcolor{blue}{// Compute Spatial Entropy $H^{\mathrm{sp}}_n$} \;
		Construct multiset $\mathcal{M}^{\mathrm{sp}}_n = \{\{(x_k, y_k\}\}_{k=1}^{N}$ \;
		Compute $p_k^{\mathrm{sp}}$ and $H^{\mathrm{sp}}_n$ using~\eqref{eq:p_sp},~\eqref{eq:spatial_entropy_multiset} \;
		\textcolor{blue}{// Compute Temporal Entropy $H^{\mathrm{tmp}}_n$} \;
		Compute the inter-arrival-time sequence
		$\{\Delta t_i\}_{i=1}^{N-1}$, where
		$\Delta t_i=t_{i+1}-t_i$ \;
		Discretize $\Delta t'_i=\Delta_b\lfloor\Delta t_i/\Delta_b+1/2\rfloor$ \;
		Construct multiset $\mathcal{M}^{\mathrm{tmp}}_n = \{\{\Delta t'_1, \dots, \Delta t'_{N-1}\}\}$ \;
		Compute  $p_k^{\mathrm{tmp}}$ and $H^{\mathrm{tmp}}_n$ using~\eqref{eq:p_tmp},~\eqref{eq:temporal_entropy_multiset}  \;
		
		\textcolor{blue}{// Compute Combined Entropy $H^\mathrm{sp\text{-}tmp}_n$} \;
		Integrate: $H^\mathrm{sp\text{-}tmp}_n = \alpha H^\mathrm{sp}_n + (1-\alpha) H^\mathrm{tmp}_n$ \;
		
		\Return $H^\mathrm{sp\text{-}tmp}_n$
	\end{algorithm}
	

	\section{An EWMA-Based Dynamic Threshold Update Algorithm}\label{sec:dynamic_threshold}
	
	The fused entropy is compared with a threshold to determine whether an attack occurs. Prior EWMA threshold methods often overlook the combination of trend and volatility in entropy-based detection~\cite{P4_2022_TDSC}. We introduce second-order exponential smoothing and fuse trend and volatility into one threshold, improving responsiveness to sudden changes while maintaining stability. Cold-start initialization, traffic adaptation, attack freezing, and smoothing constraints further stabilize updates in dynamic traffic.

	\subsection{Trend and Volatility Estimation}
	The threshold is
	\begin{equation} 
		H^\mathrm{{thd}}_n  = T_n - \kappa \cdot \sigma_n, 
		\label{eq:threshold_formula} 
	\end{equation}
	where $\kappa$ controls the volatility margin. The first and second moments and their implied deviation are
	\begin{subequations} 
		\begin{align} 
			T_n &= \beta H^\mathrm{sp\text{-}tmp}_n + (1-\beta)T_{n-1}, \label{eq:trend_T} \\
			T_n^{(2)} &= \beta (H^\mathrm{sp\text{-}tmp}_n)^2 + (1-\beta)T_{n-1}^{(2)}, \label{eq:trend_T2} \\
			\sigma_n &= \sqrt{\max\left(T_n^{(2)} - (T_n)^2,\ 0\right)} \label{eq:sigma}, 
		\end{align} 
		\label{eq:trend_and_sigma} 
	\end{subequations}
	\unskip
	\noindent where $T_n$ captures the long-term trend, $T_n^{(2)}$ is the smoothed second-order statistic, and $\sigma_n$ reflects short-term volatility. The factor $\beta\in(0,1)$ controls historical decay; larger values increase sensitivity to recent changes, whereas smaller values improve noise robustness. The maximum prevents numerical errors when the moments are nearly equal. Together, these terms capture gradual trends and short-term changes in entropy dynamics.
	
	\subsection{Adaptive Constraint Strategy}
	The following three constraints are applied to prevent abrupt threshold changes that might lead to misclassifications:
	\begin{enumerate}[label=(\roman*)] 
		\item \textit{Cold start}: the first $N_{\text{ini}}$ windows initialize $T_0$ and $T_0^{(2)}$ using their arithmetic means, compensating for the lack of historical data. No threshold is returned before initialization.
		\item \textit{Attack freezing}: if $H_n^{\mathrm{sp\text{-}tmp}}<H_{n-1}^{\mathrm{thd}}$, the EWMA states and threshold are retained so that detected attack traffic cannot contaminate subsequent updates.
		\item \textit{Adaptation and smoothing}: if an attack is misclassified as normal, a reduced threshold may cause cascading misclassifications. Therefore, for $R_n>\theta$, the threshold cannot decrease; under lower traffic, a decrease is limited to 10\% of its previous value.
	\end{enumerate}
	
	\subsection{Algorithm Implementation}
	Algorithm~\ref{alg:threshold_algorithm} has cold-start and steady-state stages. During cold start, the first $N_{\text{ini}}$ spatiotemporal entropy values are cached in $\mathcal{B}$. At the $N_{\text{ini}}$-th window, the algorithm computes $T_0$, $T_0^{(2)}$, $\sigma_0$, and the first threshold, establishing a stable basis for subsequent updates.
	
	In steady state, window $n$ is compared with the previous threshold $H_{n-1}^{\mathrm{thd}}$. If an attack is detected, the trend, second-order statistic, and threshold remain unchanged. Otherwise, $T_n$, $T_n^{(2)}$, and $\sigma_n$ are updated and a preliminary threshold is calculated. High traffic prevents a threshold decrease; under ordinary traffic, the decrease is limited to 10\%. The returned $H_n^{\mathrm{thd}}$ is used by the next window.
	
	\begin{algorithm}
		\caption{EWMA Dynamic Threshold Update Algorithm}
		\label{alg:threshold_algorithm}
		\small	\DontPrintSemicolon
		\KwData{Current spatiotemporal entropy $H^\mathrm{sp\text{-}tmp}_n$, traffic rate $R_n$, traffic rate threshold $\theta$, and previous states $T_{n-1}$, $T_{n-1}^{(2)}$, $H^\mathrm{{thd}}_{n-1}$} \;
		
		\KwResult{Dynamic threshold $H^\mathrm{{thd}}_n$} \;
		\textcolor{blue}{//  {Cold Start Stage}}\;
		\If{$n \leq N_{\text{ini}}$} {
			Cache $H^\mathrm{sp\text{-}tmp}_n$ in buffer $\mathcal{B}$ \;
			\If{$n = N_{\text{ini}}$} {
				Update $T_0 \gets \frac{1}{N_{\text{ini}}} \sum_{x \in \mathcal{B}} x$ \;
				Update $T_0^{(2)} \gets \frac{1}{N_{\text{ini}}} \sum_{x \in \mathcal{B}} x^2$ \;
				$\sigma_0 \gets \sqrt{\max(T_0^{(2)}-T_0^2,0)}$ \;
				$H^\mathrm{{thd}}_n \gets \max(T_0-\kappa\sigma_0,0)$ \;
			}
			\Return $\emptyset$
		}
		\textcolor{blue}{//  {Steady State Stage}}\;
		\eIf{$H^\mathrm{sp\text{-}tmp}_n < H^\mathrm{{thd}}_{n-1}$} {
			$T_n \gets T_{n-1}$; $T_n^{(2)} \gets T_{n-1}^{(2)}$;
			$H^\mathrm{{thd}}_n \gets H^\mathrm{{thd}}_{n-1}$ \;
		}{
			Update $T_n$, $T_n^{(2)}$, and $\sigma_n$ using~\eqref{eq:trend_T},~\eqref{eq:trend_T2}, and~\eqref{eq:sigma} \;
			$\widetilde{H}^\mathrm{{thd}}_n \gets \max(T_n-\kappa\sigma_n,0)$ \;
			\eIf{$R_n > \theta$} {
				$H^\mathrm{{thd}}_n \gets \max(\widetilde{H}^\mathrm{{thd}}_n,H^\mathrm{{thd}}_{n-1})$ \;
			}{
				$H^\mathrm{{thd}}_n \gets \max(\widetilde{H}^\mathrm{{thd}}_n,0.9H^\mathrm{{thd}}_{n-1})$ \;
			}
		}
		\Return $H^\mathrm{{thd}}_n$
	\end{algorithm}
	
	\begin{figure}[!t]
		\centering
		\includegraphics[width=3.0in]{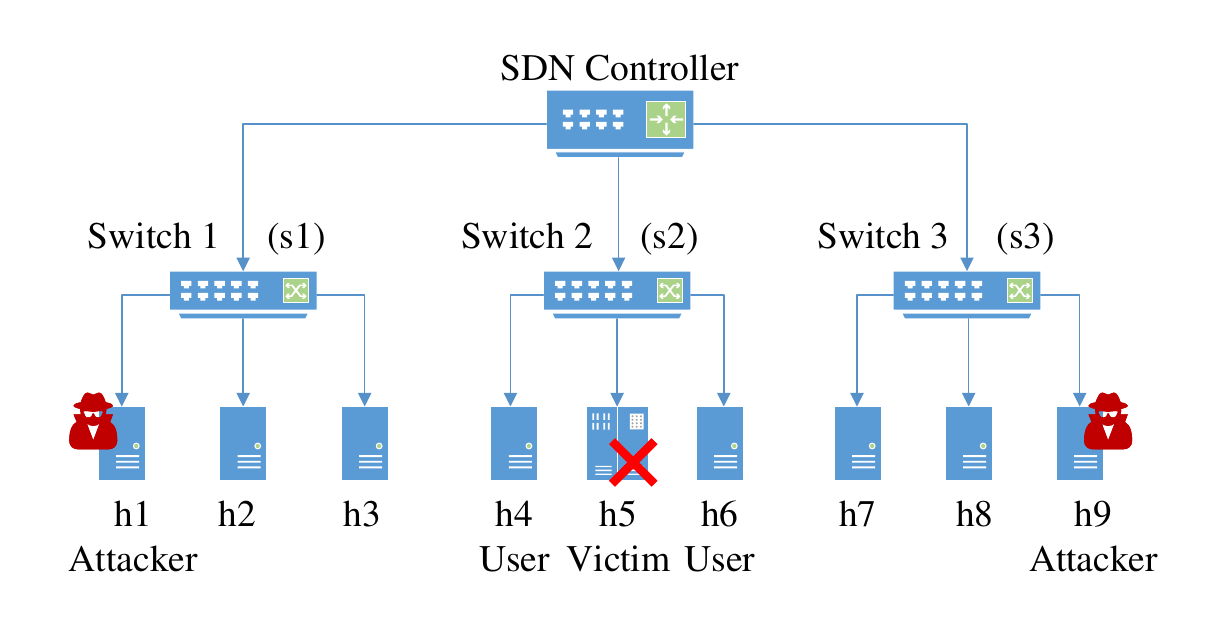}
		\caption{Three-switch SDN test topology.}
		\label{fig:topo}
		\vspace{-0.5em}
	\end{figure}
	
	\begin{figure*}[!t]
		\centering
		\subfloat[\small \fontfamily{ptm}\selectfont Spatial entropy]{
			\includegraphics[width=0.315\linewidth]{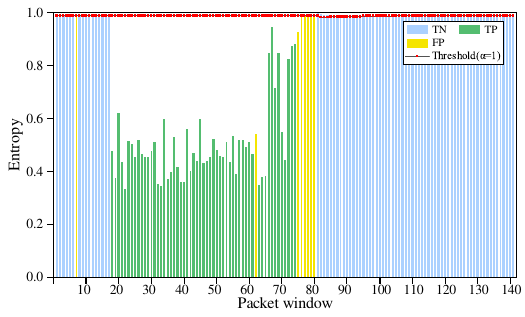}
			\label{fig:spatial_entropy}
		}
		\hfill
		\subfloat[\small \fontfamily{ptm}\selectfont Temporal entropy]{
			\includegraphics[width=0.315\linewidth]{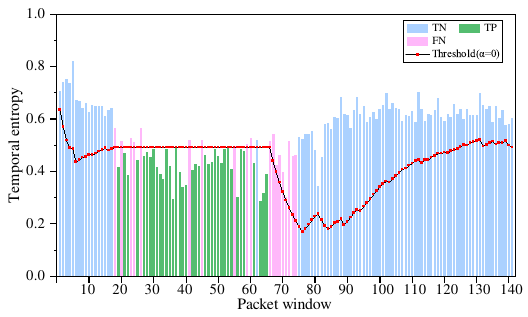}
			\label{fig:temporal_entropy}
		}
		\hfill
		\subfloat[\small \fontfamily{ptm}\selectfont Spatiotemporal entropy]{
			\includegraphics[width=0.315\linewidth]{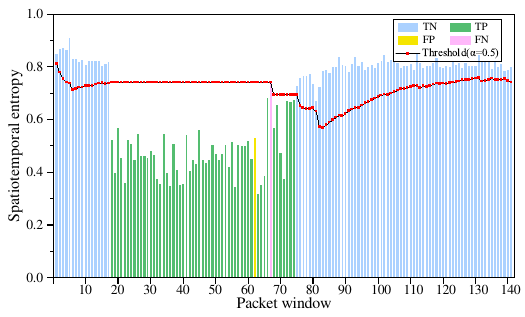}
			\label{fig:spatiotemporal_entropy}
		}
		\caption{Spatial, temporal, and fused entropy over the same traffic sequence.}
		\vspace{-1em}
		\label{fig:three_entropy}
        \vspace{-1em}
	\end{figure*}
	
	\begin{figure}[!t]
		\centering
		\includegraphics[width=2.6in]{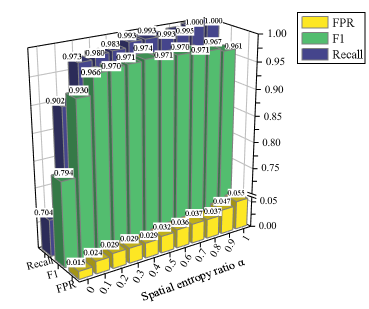}
		\caption{Sensitivity of recall, FPR, and F1-score to $\alpha$.}
		\label{fig:alpha}
	\end{figure}
	
	\section{Experiments}\label{sec:experiment}
	We evaluate entropy complementarity, adaptive thresholding, public-dataset generalization, and computational overhead.
	
	\subsection{Experimental Setup}
	\subsubsection{Simulation Environment}
	We run experiments on Ubuntu 20.04 with an Intel i9-12900K and 128 GB of memory. A Ryu 4.30 controller manages OpenFlow 1.3 switches in the Mininet 2.3.0 topology of Figure~\ref{fig:topo}; Python 3.8, Scapy, and Pyshark support implementation, traffic generation, and capture. Attackers h1 and h9 send spoofed SYN, ACK, and UDP floods toward h5, while the other hosts generate background traffic.
	
	\subsubsection{Dataset}
	A MAWI backbone trace captured between Japan and the United States on October 31, 2021, 14:00--14:15 UTC+9 provides the normal \texttt{KEYPAIR} reference~\cite{WIDE2000}. The testbed in Figure~\ref{fig:topo} generates labeled normal and spoofed SYN, ACK, and UDP flood traffic for entropy, threshold, and runtime evaluation. CICDDoS2019 provides a separate benchmark against published learning-based methods~\cite{cicddos2019developing}.
	
	\subsubsection{Evaluation Metrics}
	\label{subsec:evaluation_metrics}
	We report recall, F1-score, and false positive rate (FPR), respectively measuring detected attack windows, the precision--recall balance, and normal windows incorrectly labeled as attacks.
	
	\subsubsection{Hyperparameter Settings}
	Table~\ref{tab:hyperparameters} lists the fixed configuration selected through empirical observations and iterative testing. Window sizes, smoothing factors, and volatility coefficients were examined across multiple traffic scenarios to balance detection performance and system stability. Although these values may not be globally optimal, they provide a robust and reproducible configuration for the reported experiments.

	\begin{table}
		\centering
		\footnotesize
		\setlength{\tabcolsep}{3pt}
		\caption{Hyperparameters.}
		\label{tab:hyperparameters}
		\begin{tabular}{c|c|c}
			\hline
			Parameter & Description & Value \\\hline
			$N$ & Number of data packets in the sliding window & 400 \\
			$\Delta_b$ & Discretization granularity & 0.001 \\
			$\kappa$ & Volatility adjustment coefficient & 4 \\
			$\beta$ & Smoothing factor & 0.1 \\
			$N_{\text{ini}}$ & Number of cold start windows & 3 \\
			$\alpha$ & Weight for spatial entropy & 0.5 \\
			$\theta$ & Traffic rate threshold & 800 \\
			\hline
		\end{tabular}
	\end{table}
	
	\subsubsection{Experiment Methodology}
	Code and traffic-generation utilities are available at \url{https://github.com/johnzhang777/sted_sdn}. Figure~\ref{fig:three_entropy} shows one representative testbed run, with all entropy indicators using the same windows. Figure~\ref{fig:alpha} and Table~\ref{tab:ewma_comparison} report averages from repeated independent testbed runs. The spatial baseline reimplements~\cite{JESS2018}. The 41.74\% FPR reduction uses unrounded averages; the displayed baseline and proposed FPRs, 0.0553 and 0.0322, are rounded. Table~\ref{tab:compare on cicddos2019} separately evaluates CICDDoS2019 with random attack undersampling.
	
	\subsection{Spatiotemporal Entropy Effectiveness Analysis}\label{subsubsec:spatiotemporal_effectiveness}
	In the representative run in Figure~\ref{fig:spatial_entropy}, spatial entropy detects anomalous traffic across the attack windows, but residual anomalous traffic during Windows 75--80 causes normal traffic to be misidentified. Its recall is 1.0, confirming that all attacks are detected, while the F1-score of 0.933 and FPR of 0.094 indicate room to reduce false positives.
	
	As shown in Figure~\ref{fig:temporal_entropy}, temporal entropy produces no false positives during normal traffic phases, but misses attacks with diverse arrival intervals. Its recall and F1-score decrease to 0.661 and 0.796, respectively. Thus, temporal entropy offers stability but insufficient detection coverage when used alone.
	
	Figure~\ref{fig:spatiotemporal_entropy} combines these advantages. It removes the spatial false positives in Windows 75--80 while recovering most anomalies missed by temporal entropy, yielding 0.982 recall, a 0.982 F1-score, and a 0.012 FPR. The fused indicator therefore provides a better balance between detection coverage and precision in dynamic traffic.

	Figure~\ref{fig:alpha} shows strong average testbed performance for $\alpha\in[0.3,0.8]$. At $\alpha=1$, the detector becomes the spatial baseline, with an average FPR of 0.0553. This tolerance permits flexible deployment without precise parameter tuning, while degradation near the endpoints reflects the limitations of relying almost entirely on one entropy component. We therefore use $\alpha=0.5$ as the default.

	\subsection{Dynamic Threshold Effectiveness Analysis}\label{sec:DynamicThreshold}
	
	Table~\ref{tab:ewma_comparison} compares average testbed results for the proposed EWMA and reimplemented Cumulative Sum (CUSUM)~\cite{CUSUM}, Chebyshev~\cite{Chebyshev}, and Holt-Winters~\cite{Holt-Winters} thresholds.
	
	\begin{table}
		\centering
		\footnotesize
		\setlength{\tabcolsep}{2pt}
		\caption{Performance Comparison of EWMA with Other Dynamic Threshold Schemes.}
		\label{tab:ewma_comparison}
		\begin{tabular}{c | c c c c}
			\hline
			Metric & \makecell{Proposed\\EWMA} & \makecell{CUSUM\\ \cite{CUSUM}} & \makecell{Chebyshev\\ \cite{Chebyshev}} & \makecell{Holt-Winters\\ \cite{Holt-Winters}} \\
			\hline
			Recall & 0.9926 & 0.9540 & 0.9980 & 0.9965 \\
			F1 & 0.9737 & 0.9116 & 0.9531 & 0.9375 \\
			FPR & 0.0322 & 0.0840 & 0.0602 & 0.0814 \\
			\hline
		\end{tabular}
	\end{table}
	
	The proposed EWMA gives the best F1-score (0.9737) and lowest FPR (0.0322) with 0.9926 recall. Chebyshev and Holt-Winters achieve slightly higher recall, but their FPRs increase to 0.0602 and 0.0814, respectively; CUSUM has both lower recall and a higher FPR. Thus, the proposed EWMA provides the best overall balance between attack detection and false-alarm suppression.
	
	\subsection{Comparison with other research}\label{subsec:comparison}
	
	Table~\ref{tab:compare on cicddos2019} shows competitive performance on the randomly undersampled CICDDoS2019 sample: 0.9916 recall, a 0.9958 F1-score, and zero FPR without model training or learned feature selection. The zero-FPR result avoids unnecessary mitigation that may degrade legitimate traffic. Because preprocessing and sampling differ across studies, the table demonstrates competitiveness rather than a universal ranking.

	\begin{table}
		\centering
		\footnotesize
		\setlength{\tabcolsep}{3pt}
		\caption{Results on CICDDoS2019~\cite{cicddos2019developing}.}
		\label{tab:compare on cicddos2019}
		\begin{tabular}{c | c c c}
			\hline
			& Recall & F1 & FPR \\ \hline
			Hassan \textit{et al.}~\cite{Entropy+ML_2024_SciRep} & 0.9990 & 0.9992 & 0.0023 \\
			Wang \textit{et al.}~\cite{compare2024SciRep} & 0.9871 & 0.9834 & 0.0818 \\
			Proposed scheme & 0.9916 & 0.9958 & 0 \\ \hline
		\end{tabular}
	\end{table}

	\subsection{Algorithm Complexity and Runtime Analysis}\label{sec:Complexity}
	Computational overhead excludes packet capture, parsing, and transmission and average the overhead over ten runs with 400 packets per window. Feature extraction, \texttt{KEYPAIR} counting, inter-arrival processing, and entropy calculation require a linear scan, while fusion and threshold updates use constant arithmetic and state maintenance; the overall per-window complexity is $O(N)$. Mean latency is $3.95\pm0.026$ ms. Across all ten runs, system-wide CPU utilization increases from an idle baseline of 1\% to $11.65\pm0.74$\%. This low and stable resource consumption demonstrates that the algorithm remains lightweight under continuous multi-window processing and supports online deployment on controllers and constrained edge or IoT gateways.

	\section{Conclusions}\label{sec:conclusion}

	This study proposes a spatiotemporal entropy-based anomaly detector for SDN with an EWMA dynamic threshold. The integration combines spatial and temporal entropy with equal weights to balance sensitivity and robustness. Spatial entropy plays the primary role in perceiving attack aggregation, while temporal entropy suppresses transient false positives caused by traffic fluctuations. Testbed averages show 99.26\% recall, a 0.9737 F1-score, and a 3.2\% FPR (41.74\% below that of spatial entropy alone based on unrounded values). The dynamic threshold adapts to accepted normal traffic while preventing detected attack windows from contaminating its trend and volatility estimates. Unlike most prior approaches that adopt machine-learning classifiers, our scheme achieves a similarly high level of detection performance without requiring training or labeled data.
	
	On CICDDoS2019, the method achieves 0.9916 recall, a 0.9958 F1-score, and an FPR of 0. Its $O(N)$ complexity and 3.95 ms mean testbed core processing time support practical deployment without model training or labeled data. The protocol-agnostic design may extend to other volumetric or behavior-based anomalies. The current fixed weighting may not fully adapt to different traffic and attack characteristics; future work will investigate adaptive spatial-temporal weights and evaluate lightweight deployment in operational SDN, edge, and low-power IoT environments.

	\bibliographystyle{IEEEtran}
	\bibliography{references}
	
\end{document}